\documentstyle[epsfig]{mn}

\newif\ifAMStwofonts

\def\gtorder{\mathrel{\raise.3ex\hbox{$>$}\mkern-14mu
             \lower0.6ex\hbox{$\sim$}}}
\def\ltorder{\mathrel{\raise.3ex\hbox{$<$}\mkern-14mu
             \lower0.6ex\hbox{$\sim$}}}

\ifoldfss
  \ifCUPmtlplainloaded \else
    \NewTextAlphabet{textbfit} {cmbxti10} {}
    \NewTextAlphabet{textbfss} {cmssbx10} {}
    \NewMathAlphabet{mathbfit} {cmbxti10} {} 
    \NewMathAlphabet{mathbfss} {cmssbx10} {} 
  \fi
  \ifAMStwofonts
    \ifCUPmtlplainloaded \else
      \NewSymbolFont{upmath} {eurm10}
      \NewSymbolFont{AMSa} {msam10}
      \NewMathSymbol{\upi}     {0}{upmath}{19}
      \NewMathSymbol{\umu}     {0}{upmath}{16}
      \NewMathSymbol{\upartial}{0}{upmath}{40}
      \NewMathSymbol{\leqslant}{3}{AMSa}{36}
      \NewMathSymbol{\geqslant}{3}{AMSa}{3E}

       \let\le=\leqslant
      \let\geq=\geqslant \let\ge=\geqslant
    \fi
  \fi
\fi 

\ifnfssone
  \newmathalphabet{\mathit}
  \addtoversion{normal}{\mathit}{cmr}{m}{it}
  \addtoversion{bold}{\mathit}{cmr}{bx}{it}
  \newmathalphabet{\mathbfit} 
  \addtoversion{normal}{\mathbfit}{cmr}{bx}{it}
  \addtoversion{bold}{\mathbfit}{cmr}{bx}{it}
  \newmathalphabet{\mathbfss} 
  \addtoversion{normal}{\mathbfss}{cmss}{bx}{n}
  \addtoversion{bold}{\mathbfss}{cmss}{bx}{n}
  \ifAMStwofonts
    \ifCUPmtlplainloaded \else
      \UseAMStwoboldmath
      \makeatletter
      \new@mathgroup\upmath@group
      \define@mathgroup\mv@normal\upmath@group{eur}{m}{n}
      \define@mathgroup\mv@bold\upmath@group{eur}{b}{n}
      \edef\UPM{\hexnumber\upmath@group}
      \new@mathgroup\amsa@group
      \define@mathgroup\mv@normal\amsa@group{msa}{m}{n}
      \define@mathgroup\mv@bold\amsa@group{msa}{m}{n}
      \edef\AMSa{\hexnumber\amsa@group}
      \makeatother
      \mathchardef\upi="0\UPM19
      \mathchardef\umu="0\UPM16
      \mathchardef\upartial="0\UPM40
      \mathchardef\leqslant="3\AMSa36
      \mathchardef\geqslant="3\AMSa3E

       \let\le=\leqslant
      \let\geq=\geqslant \let\ge=\geqslant
    \fi
  \fi
\fi 

\ifnfsstwo
  \DeclareMathAlphabet{\mathbfit}{OT1}{cmr}{bx}{it}
  \SetMathAlphabet\mathbfit{bold}{OT1}{cmr}{bx}{it}
  \DeclareMathAlphabet{\mathbfss}{OT1}{cmss}{bx}{n}
  \SetMathAlphabet\mathbfss{bold}{OT1}{cmss}{bx}{n}
  \ifAMStwofonts
    \ifCUPmtlplainloaded \else
      \DeclareSymbolFont{UPM}{U}{eur}{m}{n}
      \SetSymbolFont{UPM}{bold}{U}{eur}{b}{n}
      \DeclareSymbolFont{AMSa}{U}{msa}{m}{n}
      \DeclareMathSymbol{\upi}{0}{UPM}{"19}
      \DeclareMathSymbol{\umu}{0}{UPM}{"16}
      \DeclareMathSymbol{\upartial}{0}{UPM}{"40}
      \DeclareMathSymbol{\leqslant}{3}{AMSa}{"36}
      \DeclareMathSymbol{\geqslant}{3}{AMSa}{"3E}

       \let\le=\leqslant
      \let\geq=\geqslant \let\ge=\geqslant
    \fi
  \fi
\fi 

\ifCUPmtlplainloaded \else
  \ifAMStwofonts \else 
    \def\upi{\pi}
    \def\umu{\mu}
    \def\upartial{\partial}
  \fi
\fi

\title[Supernovae in Deep Hubble Space Telescope Galaxy Cluster Fields: 
Cluster Rates and Field Counts] {Supernovae in Deep Hubble Space Telescope\\
 Galaxy Cluster Fields: Cluster Rates and Field Counts$^{1}$}

\date{Accepted - .
      Received - ;}

\author[A. Gal-Yam et al.]{Avishay Gal-Yam$^{2,3}$, 
Dan Maoz$^{2,4}$,
and Keren Sharon$^{2}$\\
$^{1}$ Based on observations made with the {\it Hubble Space
Telescope}, which is operated by AURA, Inc., under NASA contract NAS5-26555.\\
$^{2}$ School of Physics \& Astronomy and Wise Observatory, Tel Aviv University, Tel Aviv 69978, Israel; avishay@wise.tau.ac.il \\
$^{3}$ Colton Fellow. \\
$^{4}$ Department of Astronomy, Columbia University,
550 W. 120th St., New York, NY 10027, USA; dani@wise.tau.ac.il}

\begin{document}

\maketitle


\begin{abstract}

We have searched for high-redshift supernova (SN)
candidates in multiple deep Hubble Space Telescope ({\it HST}) 
archival images 
of nine galaxy-cluster fields. We detect six apparent SNe, with 
$21.6 \le I_{814} \le 28.4$ mag. There is roughly 1 SN per deep 
($I_{814} > 26$~mag), doubly-imaged,
WFPC2 cluster field. Two SNe are associated with cluster galaxies
(at redshifts $z=0.18$ and $z=0.83$), three 
are probably in galaxies not in the clusters (at $z=0.49$, $z=0.60$, 
and $z=0.98$), and one is at unknown $z$. 
After accounting for observational efficiencies
and uncertainties (statistical and systematic) we derive the rate of 
type-Ia SNe within the projected central $500h^{-1}_{50}$ kpc of rich clusters:
$R=0.20^{+0.84}_{-0.19} h_{50}^{2}$ SNu in $0.18\le z \le0.37$ clusters, and 
$R=0.41^{+1.23}_{-0.39}h_{50}^{2}$ SNu in clusters at $0.83\le z \le 1.27$
(95 per cent confidence interval; 1 SNu $\equiv$ 1 SN 
century$^{-1}$ per $10^{10} L_{B \odot})$.
Combining the two redshift bins, the mean rate is
$R_{\bar z=0.41} = 0.30^{+0.58}_{-0.28} h_{50}^{2}$ SNu. 
The upper bounds argue against SNe-Ia
being the dominant source of the large iron mass
observed in the intra-cluster medium.
We also compare our observed counts of field SNe (i.e., non-cluster SNe
of all types) to recent model predictions.
The observed field count is $N\le 1$ SN with $I_{814}\le 26$ mag, 
 and $1\le N \le 3$ SNe with $I_{814}\le 27$ mag. These counts are
about two times lower than some of the predictions. Since the counts at these
magnitudes are likely dominated by type-II SNe, 
our observations may 
suggest obscuration of distant SNe-II, or a SN-II luminosity
distribution devoid of a large high-luminosity tail. 
\end{abstract}


\section{Introduction}

The search for distant supernovae (SNe) 
has been revolutionised in the last few years by
two cosmology-oriented groups, the Supernova Cosmology Project (SCP;
Perlmutter et al. 1997), and the High$-z$ SN Search Team (Schmidt et al. 
1998). Both groups have discovered hundreds of distant SNe with redshifts 
up to $z=1$, and beyond. Analysis of these data shows evidence 
for an accelerating Universe, possibly driven by a positive 
cosmological constant (Riess et al. 1998; Perlmutter et al. 1999).
Two likely high$-z$ SNe have also been discovered in a repeated observation
with the  Hubble 
Space Telescope ({\it HST}) of the Hubble Deep Field 
(HDF; Gilliland, Nugent, \& Phillips 1999). One of these SNe has been
shown by Riess et al. (2001) to be a type-Ia at $z=1.7$, at an epoch
before the transition from deceleration to acceleration. 
A third possible 
SN candidate was detected in the original HDF data through the analysis of 
the time dependent flux variation of objects in this field (Mannucci \& 
Ferrara 1999). These observations demonstrate that deep {\it HST} 
observations are an efficient way to find high$-z$ SNe, with the 
high angular
 resolution and low sky background compensating for the small field 
of view.

Apart from the use of type-Ia SNe as standard candles, it is important
to find and characterise distant SNe of all types. SN rates as a function
of redshift probe the star formation history of the Universe, the physical
mechanisms leading to SN-Ia explosions, and the cosmological parameters
(Jorgensen et al. 1997; Sadat et al. 1998; Ruiz-Lapuente \& Canal 1998; 
Madau, Della Valle, \& Panagia 1998;
Yungelson \& Livio 2000). Furthermore, even in the
absence of redshift information, SN number counts as a function of 
limiting magnitude can provide valuable constraints (Dahl\'en \& Fransson 1999,
hereafter DF; Sullivan et al. 2000, hereafter S2000).
  
There are incentives to search for high$-z$ SNe
not only in ``blank'' fields, but also in the
fields of rich galaxy clusters. The lensing 
magnification of background sources by the cluster potentials
enables detection of distant and intrinsically dim sources, that would be 
too faint to study otherwise. Indeed, 
the natural telescopes provided by rich clusters have been used successfully 
for deep studies of high$-z$ galaxies in the X-ray (Crawford et al. 2001),
optical (e.g., Frye \& 
Broadhurst 1998, Pettini et al. 2000), IR (e.g., Lemonon et al. 1998), and 
submm (Smail et al. 1998) regimes. Following the same reasoning, 
S2000 have considered the prospects of
searching for high$-z$ SNe lensed by galaxy clusters. 
The most distant SNe could be found, with 
implications both for the
cosmological applications and for our understanding of the nature of SNe.

Lensed high$-z$ SNe behind
galaxy clusters could also probe the properties of the cluster lenses.
For instance, a magnified type-Ia SN 
would provide a direct measurement of the cluster magnification,
lifting the so called ``mass-sheet degeneracy'' encountered when deducing 
cluster masses from weak-lensing shear maps (Kolatt \& Bartelmann 1998). A 
SN in a host galaxy that is lensed by a cluster into a giant arc may be
multiply imaged into three images with time delays of weeks to months 
(Kovner \& Paczynski 1988). SNe  in galaxies that are nearly lensed
into optical Einstein rings, such as 
the CL0024+16 arc system (Tyson, Kochanski, \& dell'Antonio 1998),
may also have short time lags (months to years) between images of the SNe
in distinct galaxy images, providing
constraints on the cluster mass distribution and on cosmological parameters.

Finally, knowledge of the rate of SNe {\it in} the galaxy clusters,
as a function of SN type and cluster redshift,
could illuminate several puzzles concerning the 
intra-cluster medium (ICM). One such puzzle is the total observed 
mass of iron, which is at least several times 
larger than that expected from SNe, based on the present-day stellar masses,
and assuming a standard stellar initial mass function (IMF; 
e.g., Brighenti and Mathews 1998; Loewenstein 2000). Another is the 
energy budget of the ICM gas and the ``entropy floor'' observed in clusters,
which suggest a non-gravitational energy source to the ICM, again 
several times larger than the expected energy input from SNe
(e.g., Lloyd-Davies, Ponman, \& Cannon 2000; Tozzi \& Norman 2001; 
Brighenti \& Mathews 2001). Proposed 
solutions to these problems have included an IMF skewed toward high-mass
stars (so that a large number of iron-enriching core-collapse SNe 
are produced per
present-day unit stellar luminosity), or a dominant role for SNe-Ia in
the ICM enrichment. The latter option would require early-type galaxies 
in clusters to have SN-Ia rates much higher than observed in nearby 
field ellipticals. Constraints on these scenarios can be obtained by 
directly measuring the SN rate in clusters at various redshifts.   

Observationally, SN searches in galaxy clusters were pioneered in the
late 1980's by Norgaard-Nielsen et al. (1989), resulting in the first
detection of a $z=0.31$ SN in the galaxy cluster AC118. More recently,
low-redshift cluster samples have been monitored for SNe by the Mount
Stromlo Abell Cluster SN Search (Reiss et al. 1998) and by the Wise 
Observatory Optical Transients Search (Gal-Yam \& Maoz 1999, 2001). 
To extend these efforts to higher$-z$ clusters, 
we have carried out a search for SNe in duplicate deep cluster images
in the {\it HST} archive. In this paper, we describe our discovery of
six SNe in this data set. We then use these SNe to derive the SN-Ia rate
in clusters and to test the theoretical
predictions for the non-cluster SN number counts in such fields.
Where applicable, we assume a flat cosmology with parameters
$\Omega_{m}=0.3$ and $\Omega_{\Lambda}=0.7$, and use $h_{50}$ to denote
the Hubble parameter in units of 50 km s$^{-1}$ Mpc$^{-1}$.

\section{Archival target selection and analysis}

We have searched the {\it HST} archive for duplicate WFPC2 observations 
of galaxy cluster fields. Clusters were identified as such by the target 
name, or, when the names were not revealing, by searching the NED 
database for known clusters near the field coordinates. The details of our 
final cluster sample are given in Table 1. For two WFPC2 observations to 
be considered duplicates, we required that they had been obtained through the 
same filter, and that the angular distance between their field centres was 
less than $75$ arcsec. The requirement for identical filters is due to the need 
for accurate image subtraction when searching for faint transients.
The angular separation criterion is driven by the peculiar
shape of the WFPC2 field, coupled with the random {\it HST} orientations,
which can
result in observations that are nominally duplicate, but have little
overlap area.
  
\begin{table*}
\caption {Galaxy Cluster Fields with Duplicate {\it HST}
Observations} \label {cluster table}
\vspace{0.2cm}
\begin{centering}
\begin{minipage}{140mm}
\begin{tabular}{ccccrrcccc}
\hline
Cluster & $z$  & Epoch  & Exposure & Sub-  &Filter& Program& Overlap Area$^{a}$ & Number of \\
        &      & (UT Start)  & [ks] & frames& &ID     &  [arcmin$^2$] & Apparent SNe \\
\hline
Abell 2218    & 0.18  & Mar 11 1999 & 8.4  & 12 & F606W & 7349 & 4.69  & 1 \\ 
              &       & Jan 13 2000 & 10.0 & 10 & F606W & 8500 & 4.69  & - \\
Abell 1689    & 0.18  & Jun 27 1996 & 2.0  &  2 & F814W & 6004 & 0  & -    \\
              &       & Jun 19 1997 & 4.0  &  4 & F814W & 6004 & 4.69  & - \\
              &       & Jul 07 1997 & 4.6  &  4 & F814W & 5993 & 1.56  & - \\
              &       & Feb 24 2000 & 4.0  &  4 & F814W & 8546 & 1.56  & - \\
AC114         & 0.31  & Jan 06 1996 & 16.6 &  6 & F702W & 5935 & 1.28  & - \\ 
              &       & Oct 26 1997 & 15.6 &  6 & F702W & 7201 & 1.28  & - \\
MS1512.4+3647 & 0.37  & Mar 19 1996 & 6.3  &  9 & F555W & 6003 & 2.30  & - \\
              &       & Jun 20 1997 & 10.4 &  8 & F555W & 6832 & 2.30  & - \\
MS2053.7$-$0449 & 0.58  & Oct 23 1995 & 2.6  &  2 & F814W & 5991 & 0   & - \\
              &       & Sep 25 1998 & 3.2  &  3 & F814W & 6745 & 4.42  & - \\
MS1054.4$-$0321 & 0.83  & Mar 13 1996 & 15.6 &  6 & F814W & 5987 & 4.69 & 3 \\
              &       & May 29 1998 & 6.6  &  6 & F814W & 7372 & 4.69  & - \\
CL1604+4304   & 0.89  & Feb 25 1994 & 32.0 & 16 & F814W & 5234 & 3.97  & 1 \\
              &       & Feb 28 1995 & 32.0 & 16 & F814W & 5234 & 3.97  & 1 \\
RXJ0848.6+4453& 1.27  & Mar 01 1999 & 13.9 &  5 & F814W & 6812 & 4.69  & - \\
              &       & Apr 15 1999 & 13.9 &  5 & F814W & 6812 & 4.69  & - \\ 
CL1645+46     &$>2.8$ & May  7 1997 & 6.0  & 12 & F814W & 6598 & 3.42  & - \\
              &       & Sep 30 1999 & 5.3  &  4 & F814W & 7342 & 3.42  & - \\
\hline
\end{tabular}

Notes:\\ 
$^a$ The overlap area is set to zero for data sets with fewer then
three sub-exposures. These data sets are used only as references 
for searching for SNe at other epochs (\S~3.3.3).\\

\end{minipage}
\end{centering}
\end{table*}

To optimise the search for SNe, especially at high $z$, where cosmological 
time-dilation is important, we required a minimal time separation
between data sets of 30 days. The implications of this criterion are 
accounted for in the analysis of our results. 
Finally, we have limited our search to deep observations, with total
exposure times of at least 2000~s in one of the broad-band {\it HST}
filters. In order to facilitate reasonable rejection of cosmic rays and 
hot pixels, we required that at least one of the duplicate data sets consist 
of at least three sub-exposures. When both duplicate data sets had three 
or more sub-exposures, we searched for candidate variable objects in 
both. In cases where only one of the data sets was suitably split, the other 
set was used only as a reference, since the possibility of cosmic ray
events mimicking transient objects could not be ruled out.
We have found nine galaxy cluster fields with observations
in the {\it HST} archive satisfying these criteria. 
The overlap between images from
different epochs is partial, so the total effective area useful for a
bi-directional
search is 30.8 arcmin$^{2}$, or 6.6 times the total area of WFPC2,
excluding the PC1 CCD (see Table 1).  

Sets of WFPC2 images at a given epoch were combined using 
software based mainly on 
the {\it DITHER} package (Fruchter \& Hook, 1997) within IRAF 
\footnotemark[1].  
\footnotetext[1]{IRAF (Image Reduction and Analysis Facility) is distributed
by the National Optical Astronomy Observatories, which are operated
by AURA, Inc., under cooperative agreement with the 
National Science Foundation.}
A flexible, high-fidelity algorithm was used to reject cosmic-rays and 
hot/cold pixels, even in stacks with as few as three sub-exposures, 
allowing their use in this project. The PC1-CCD data were
not used in the analysis, due to the lower sensitivity, the relatively
small addition of area, and the need to resample the data in order
to compare to WF-CCD data from other epochs.

The combined images of a cluster from two available epochs were
registered, using a general geometrical transformation, calculated from
the positions of $\sim20$ compact galaxies or stars, visible at both epochs.
The images were scaled according to their exposure times and subtracted,
forming a difference image. The stable {\it HST} point-spread function 
(PSF) allows for good cancellation of almost all non-variable objects,
as shown in an example in Fig. 1. 
Some low-level residuals remain in the difference images 
near bright stars and at the nuclei of the brightest galaxies. These 
residuals are unlikely to mimic real objects, since their suspect origin 
is evident from the images. However, the residuals near galactic nuclei 
limit the detection of nuclear variability in these galaxies, either 
SNe close to the nucleus or active galactic nuclei (AGN). 
This is taken into account 
by our detection efficiency simulations, described in
 \S~3.3.2.

\begin{figure}
\centerline{\epsfxsize=85mm\epsfbox{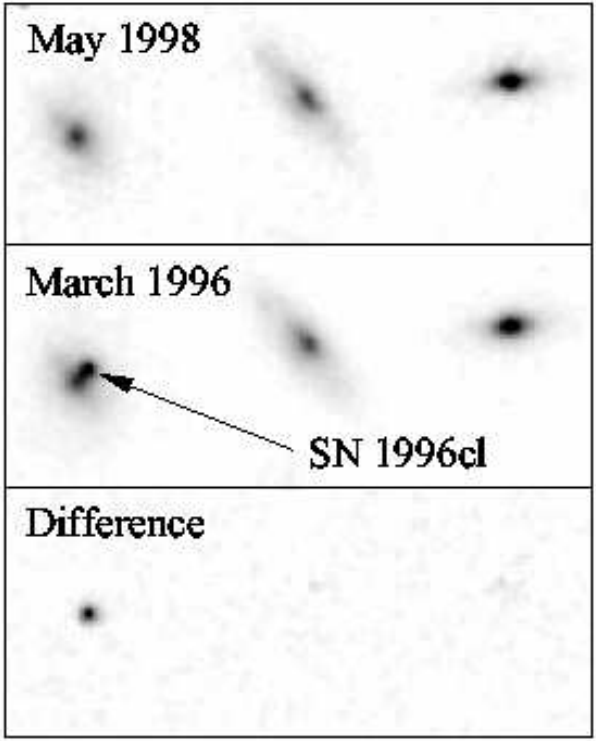}}
\caption{Example of image subtraction; SN 1996cl, in a galaxy  in
 cluster MS1054.4$-$0321, in
 a $8\farcs5 \times 3\farcs6$ section of the difference image (lower panel) 
created by subtracting the data  obtained
in May 1998 (upper panel) from the data of the same field taken during 
March 1996 (middle panel). Note the clean subtraction of the galaxies,
resulting from the stable {\it HST} point-spread function.
}
\end{figure}

The difference images were methodically scanned by eye, and all residual 
objects checked and classified. After the
elimination of obvious subtraction artefacts, the location of each residual
object was checked in each of the sub-exposures from that epoch. For a
candidate to be considered secure, we required that  it 
 have a stellar PSF, and
appear in 
at least three independent subsets of the data, each consisting of
the sum of individual exposures
in which its location is not obviously affected 
by cosmic rays, with non-variable flux levels, to within errors.
This criterion eliminates residual events 
consisting of cosmic rays that are coincident in several sub-exposures,
as well as chip defects and rare, high amplitude, background
fluctuations, since in these cases most of the candidate's flux in the final
combined image comes from only some of the available sub-exposures.

Care was taken not to reject real events affected by cosmic ray hits, even
in data sets with few sub-exposures. Real events could, in principle, be
lost in three-image data sets if cosmic rays happened to hit the object's
location on two of the three sub-exposures. However, given
that only 1-2 per cent of the pixels in each image are affected
by cosmic rays, this has a probability
of $\sim10^{-4}$, and much lower in data sets with more than three
sub-exposures.  
Candidates that passed these criteria were considered real transient
or variable sources, most likely high$-z$ SNe (see \S3.2). 
In practice, all but one of the
transient sources we found can be identified on each individual exposure, 
and not only in summed subsets. Figure 2 shows an example.

\begin{figure}
\centerline{\epsfxsize=85mm\epsfbox{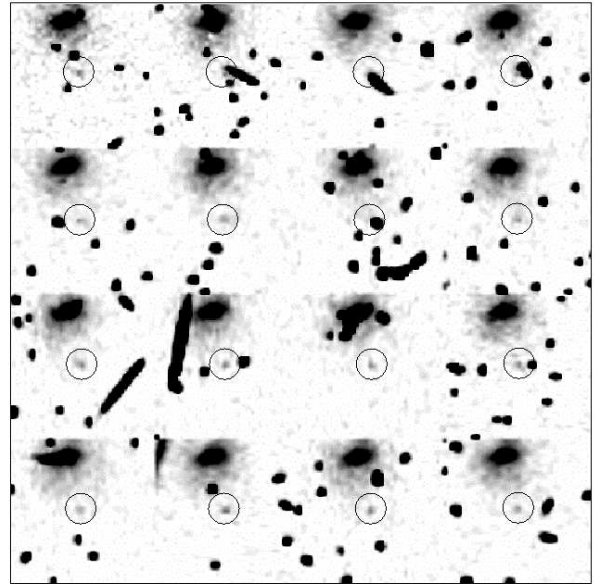}}
\caption{Example of candidate verification; 
$2\farcs1 \times 2\farcs1$ sections of the 
individual WFPC2 exposures 
at the location of SN 1995bf in CL1604+4304. 
The apparent SN is visible in all 
sub-exposures where the location is not affected by cosmic-ray
 hits, with a 
stellar PSF and constant flux, to within the expected errors.
}
\end{figure}

The flux limit at each epoch
depends on the length of the exposures and the filter used, and 
therefore varies between our sample clusters, but in all sets the limit 
for $>50$ per cent detection is better than $I_{814}=26$ mag\footnotemark[2].  
\footnotetext[2]{We report our observations using the AB magnitude 
system, in which the magnitude 
in a band $b$ centered at wavelength $\lambda$ [nm] is 
$b_{\lambda}= -48.6 -2.5 \times \log f_{\nu}$, where $f_{\nu}$
is the flux density at frequency $\nu$ in 
erg s$^{-1}$ cm$^{-2}$ Hz$^{-1}$.}
 
\section{Results}

\subsection{Apparent supernovae}

We have detected six apparent SNe in the cluster fields we have searched, 
including the rediscovery of SN 1996cl, which was
previously known. For each SN, 
Table 2 lists the details, and Fig. 3 shows
a section of the images at two epochs.  Among the six  
SNe, five are projected within 2.5 half-light radii of galaxies that
are likely their hosts, while the sixth has no obvious host
(but see $\S~3.1.2$). 
We have been able to determine
the redshift and types of the host galaxies of five events from the 
literature (see below). 
As noted in proof by S2000, the object they 
previously considered a SN candidate, near the core of AC114, is 
not variable. Indeed, Campusano et al. (2001) show that
this object is likely one of the images of a 
multiply-lensed background galaxy (marked A4 in their Fig.~1).    
We briefly discuss each SN below.

\begin{figure*}
\epsfig{figure=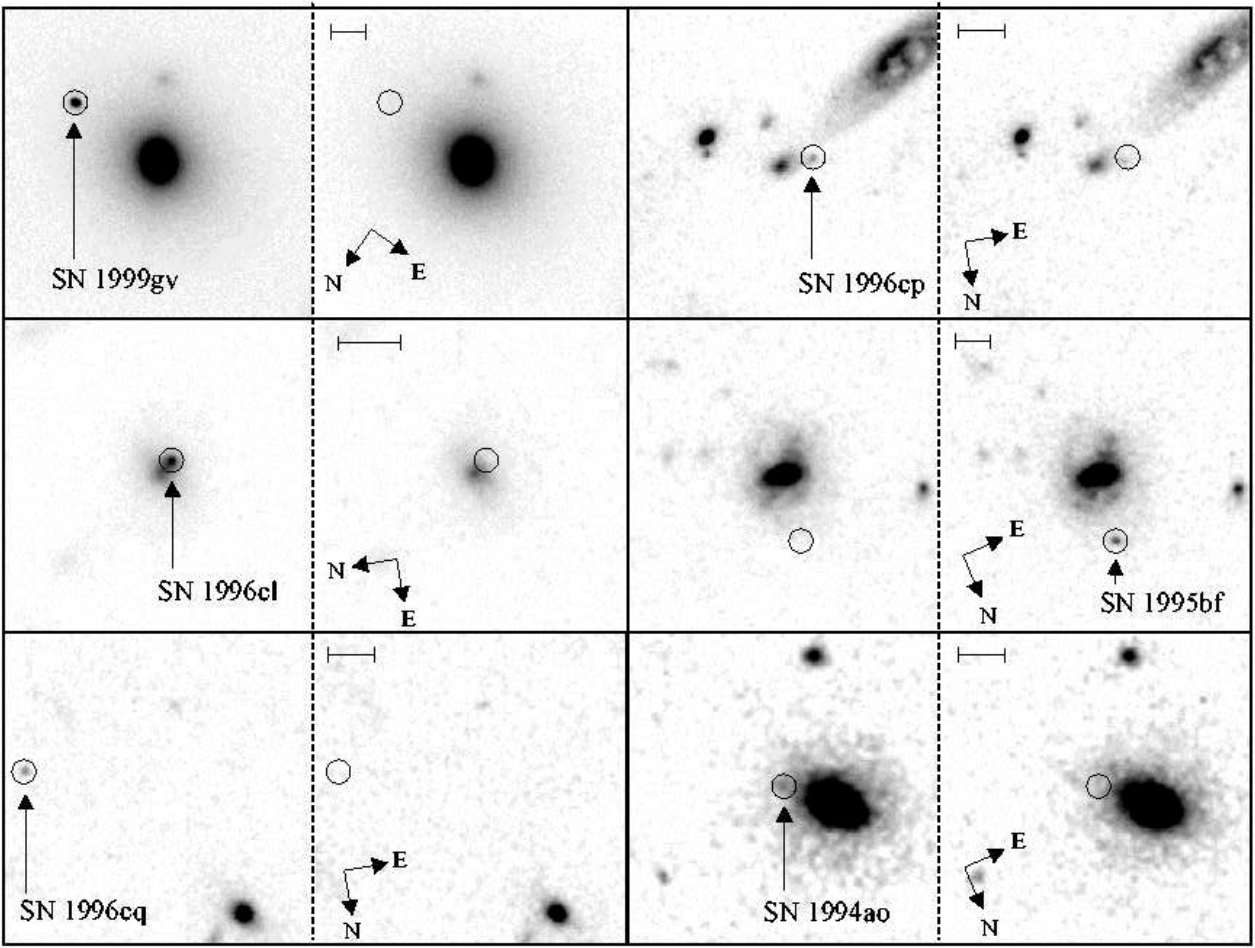,angle=0,width=170mm}
\caption{Sections of the images, at two epochs, for each 
of the six apparent SNe. 
The scales shown in the upper-left-hand corners correspond
to $1''$. 
}
\end{figure*}

\begin{table*}
\caption {Apparent Supernovae} \label {sn table}
\vspace{0.2cm}
\begin{centering}
\begin{minipage}{140mm}
\begin{tabular}{cccccccc}
\hline
SN & Cluster & Magnitude &  $\alpha$ & $\delta$ & Offset  & Host \\ 
   & Field   &           &  (2000)   & (2000)   & from Host & $z$ \\    
\hline
1999gv & Abell 2218    & $V_{606}=21.64 \pm 0.03$ & $16^h36^m02.2^s$ & $+66^{\circ}12'34"$ & $0\farcs8$ S $2\farcs8$ W & 0.175\\
1996cl & MS1054.4$-$0321 & $I_{814}=23.71 \pm 0.04$ & $10^h56^m59.1^s$ & $-3^{\circ}37'36"$  & $0\farcs1$ S $0\farcs05$ W & 0.827\\
1996cq & MS1054.4$-$0321 & $I_{814}=25.61 \pm 0.05$ & $10^h57^m02.2^s$  & $-3^{\circ}37'34"$  & - &  ?   \\ 
1996cp & MS1054.4$-$0321 & $I_{814}=26.7 \pm 0.2$ & $10^h57^m03.5^s$  & $-3^{\circ}37'41"$  & $0\farcs9$ E & 0.596$^a$   \\    
1995bf & CL1604+4304   & $I_{814}=26.14 \pm 0.06$ & $16^h04^m20.4^s$ & $+43^{\circ}04'44"$ & $2\farcs9$ N $0\farcs6$ W & 0.985\\
1994ao & CL1604+4304   & $I_{814}=28.4 \pm 0.5$ & $16^h04^m23.3^s$ & $+43^{\circ}03'56"$ & $1\farcs4$ S $1\farcs1$ W & 0.496\\
\hline
\end{tabular}

Notes:\\ 
$^a$ Other host galaxies possible, see text.\\

\end{minipage}
\end{centering}
\end{table*}

\subsubsection{SN 1999gv in the field of Abell 2218}

During March 1999, a large mosaic (22 positions) of WFPC2 images,
covering the entire field of Abell 2218, was obtained through the
F606W filter. The main goal of this program (PI Squires) was 
to create weak lensing
shear maps, and to compare the resulting mass distribution to
that derived via strong lensing, X-rays, and the Sunyaev--Zeldovich
effect. In January 2000, following the  
Space Shuttle servicing mission 3a, the cluster was re-imaged 
through the F450W, F606W, and F814W filters as an
``early release'' observation, intended to demonstrate the telescope
performance. Comparison between the
8.4-ks F606W image from March 1999 and the 10.0-ks 
F606W image from January 2000, reveals a bright unresolved object  
in the 1999 exposure (Gal-Yam \& Maoz 2000; see Fig. 3).
The host galaxy 
is catalogued as No. 149 in Le Borgne et al. (1992), where it is classified
as an elliptical galaxy  
with $B = 19.9$ mag and redshift $z = 0.1753$.
In view of the brightness of this SN, and its association with an 
elliptical cluster member, it is most likely a type-Ia SN in the cluster.
Assuming a typical SN-Ia, the apparent magnitude suggests
 a SN  age of $\sim 25$ days past maximum light.  
  
\subsubsection{SNe 1996cl, 1996cp, and 1996cq in the field of 
MS1054.4$-$0321}

The core of the distant galaxy cluster MS1054.4$-$0321 $(z=0.83)$ was
observed with the F814W filter during March 1996, in
 a program to study distant, X-ray selected clusters (PI
Donahue). It was re-observed in May 1998, as part of a 
shallower, wide field mosaic (PI Franx), designed for
a morphological large-scale study. A comparison 
of the images from the two epochs reveals 
three apparent SNe (Gal-Yam, Sharon, \& Maoz 2001).
    
The brightest of the three is SN 1996cl, 
originally discovered by the SCP in ground-based data
obtained on March 18, 1996 (Perlmutter et al. 1996), and 
spectroscopically confirmed 
as a type-Ia SN at the cluster redshift.
In the {\it HST} observation, 
obtained 5 days prior to that of Perlmutter et al.,
the SN is detected with $I_{814}=23.71~$ mag, 
the magnitude expected for SN-Ia some
10 days ($\sim5.5$ days in the SN rest frame) 
before peak magnitude.

 A second transient object in this field, SN 1996cq,
 is detected at $I_{814}=25.61~$ mag, without any obvious host galaxy. 
This object is superposed on a faint fuzz near the image detection limit 
that may be a low-surface-brightness galaxy associated with it. 

The third transient source is SN 1996cp, with
$I_{814}=26.7~$ mag. This object is superposed on a small knot
of emission, $0.9$ arcsec due east from the nucleus of a 
blue, late type galaxy with $z=0.60$ (Milvang-Jensen et al. 2002).
However, it could also be associated with a larger spiral galaxy 
that is found $4.1$ arcsec to the south-east,
catalogued as No. 1403 by van Dokkum et al. (2000), who classify 
it as a cluster member with $z=0.81$. Since the emission knot
which hosts SN 1996cp is closer to the $z=0.60$ galaxy and has
similar $I_{814}-V_{606}$ colours, we adopt a redshift of $z=0.60$
for this SN. However, we cannot at this time exclude the possibility
that the SN is associated with the nearby cluster galaxy (No. 1403)
or that the emission knot is a separate galaxy
altogether. Derivation of photometric redshifts for the
emission knot hosting SN 1996cp from multi-color images of this cluster 
may shed more light on this issue.

\subsubsection{SNe 1994ao and 1995bf in the field of CL1604+4304}

The cluster CL1604+4304 ($z=0.89$) was the target of a deep 
observation in 1994 (32~ks through the F814W filter; PI Westphal). 
However, a pointing error placed the target 
 $\sim20$ arcsec from the designated position. The cluster was therefore 
re-imaged one year later, again for 32 ks.
Subtraction of these two deep images reveals two SNe 
(Gal-Yam et al. 2001). The first,
SN 1995bf, with $I_{814}=26.14~$ mag, is found 
near a putative host, catalogued by Lubin et al. (1998) as No. 019, 
an Sb galaxy with $z=0.985$, i.e., behind the cluster.

The limiting magnitude for SN detection in these particular images, 
derived from our detection-efficiency simulations (see \S3.3.2), 
is $I_{814}=29$ mag, if the requirement that the SN appear in individual sets
of 
sub-exposures is relaxed, to the effect that candidates need
be detected 
with a stellar PSF and constant flux in only two
independent subsets of the data. 
This analysis reveals a second, very faint, apparent SN in this cluster. 
SN~1994ao has $I_{814}=28.4~$ mag,
near the detection limit,
 and is the faintest SN detected to date ($\sim1.5$ mag fainter than 
SN 1997ff; Gilliland et al. 1999; Riess et al. 2001). This object is near an S0
 galaxy with $z=0.496$, in the cluster foreground  
(No. 028 of Lubin et al. 1998). If the event is indeed associated with 
this host, it could imply it is a SN Ia, as
core-collapse SNe (types II, Ib, Ic) are generally not found in S0 
galaxies. However, the absence of core-collapse SNe
in S0 galaxies has only been established in low redshift $(z\le0.1)$ 
environments. If this
is indeed a type-Ia SN, then the low luminosity  implies it
is many months past maximum.
We report this event as an example of the capability of {\it HST} to find
very faint SNe, but do not include it in our quantitative analysis
of the results, below.

\subsection{Supernova alternatives}

Apart from SN 1996cl, all the apparent SNe we discuss lack spectroscopic
confirmation.
We now consider alternative explanations, other than SNe,
for the observed objects, and argue that all are unlikely.
 
{\bf Solar system objects --} Small bodies in the Solar System
(e.g., Kuiper Belt 
objects -- KBOs; Trujillo, Jewitt \& Luu 2001, and references therein)
can potentially mimic transients. To pose as one of our apparent 
SNe, such an object must be faint ($I_{814} > 22$~mag) and slowly moving.
However, the combination of relatively long observing periods (hours to 
days) with the {\it HST} resolution allows us to set firm upper 
limits on the proper motions of the detected transient events. 
For the relevant data set 
with the shortest observing period (MS1054.4$-$0321), the upper limit 
on the motion of the transients is $\mu \le 0.12''$day$^{-1}$.
For all three fields in which we have found SN candidates, the expected 
parallaxes of a KBO at 50~AU, in the field directions and at the times 
of observation, as a result of the Earth's orbital motion, 
are $>70''$day$^{-1}$. This is much larger than
our upper limit. We note that the proper motion of a KBO at several
tens of AU from the Sun cannot be large enough to offset the effect 
of the Earth's parallax, so that KBOs, regardless of their exact orbital
parameters, cannot account for any of the events we discuss.
  
{\bf Variable Galactic stars --} 
The fainter ($I_{814} \sim 26$~mag) among the  transient events we have discovered
are generally too faint to be plausible  
variable Galactic objects.  Such objects would need to vary by 
at least $\sim1$ mag in order to 
fall below the detection limit on one of the epochs.
A chromospherically active $I=27$~mag M4 dwarf 
(which, say, flared into our detection range), would
be at a distance of 40 kpc, far out in the low-stellar-density halo,
and hence highly unlikely to turn up in such a small field. 
Cataclysmic variables in quiescence will always be brighter
at minimum light than a single M-dwarf. Our
only SN candidate brighter than $I_{814} = 25.5$ that lacks spectroscopic
confirmation, SN 1999gv in Abell 2218, could be a  
bright flare from a Galactic G- or K-dwarf. However, the
lack of any evidence for such flares at this position 
in numerous previous images of
Abell 2218, obtained over several years through various telescopes
(e.g., Gal-Yam \& Maoz 2000), makes this possibility unlikely.

{\bf Variable stars in other galaxies --} 
If the transient events we have found are associated with the galaxies near
to which they are projected, then they are generally too luminous to be
anything but SNe.
Of the six events
we discuss, the redshifts of the probable hosts of five events are known 
(Table 2).
The most luminous novae have
absolute magnitudes $M_I=-11$ (Della Valle 1991; Warner 1995) and thus 
would be far too faint to be 
detected at the redshifts of these galaxies, falling below $I\sim28.5$ mag
at $z\sim0.17$. The host galaxy 
of SN 1996cq is only marginally detected in the deep {\it HST} image. 
If this event is a nova, the implied luminosity of the host becomes
implausibly low, $M_{I}>-8.7$~mag.

{\bf Active galactic nuclei --} 
None of the events we discuss are located in the 
nuclei of the putative host galaxies, and hence are not
Seyfert-like AGN. The candidates could, in principle, be more distant,
highly variable quasars in the nuclei of undetected galaxies,
projected by chance within $<2.5$ half-light radii of the 
 galaxies that are seen in five of six cases. By performing aperture 
photometry for all the galaxies in each image, we have determined the
galaxies' half-light radii. We find that, even in the most crowded fields,
$<5$ per cent of the total area we have searched
is within $2.5$ half-light radii of a galaxy.
For SN 1996cq, whose candidate host is barely detected, the quasar option
cannot be ruled out as firmly.

{\bf Gamma-ray burst afterglows --} Recent observations of
gamma-ray burst optical afterglows 
suggest that
these explosions are not isotropic
(e.g., Frail et al. 2001, and references therein).
If the gamma rays are more beamed than the optical emission,
there may be many optical afterglows that have 
no corresponding gamma-ray burst (Rhoads 1997). Such events could, 
in principle, explain some of the transients we have discovered. 
However, recent work
by Dalal, Griest, \& Pruet (2001) shows that even a continuous 
monitoring campaign reaching a limiting magnitude of $R = 27$
is expected to yield $<0.15$ events deg$^{-2}$ yr$^{-1}$.
Our search has visited  a total area $< 0.01$ deg$^{2}$, only
twice.
     
We conclude that most or all of the events we have discovered 
are likely bona-fide SNe. Nevertheless, in our derivation of 
cluster SN-Ia rates and field counts, the estimated uncertainties account 
for the more ambiguous identifications of several of the candidates.
In our comparisons of field counts to some theoretical predictions (\S3.3.3),
if any of the events were not SNe, the discrepancy between 
observations and some models would only grow.   

\subsection{Field supernova counts}
\subsubsection{Observed counts}

Of the six apparent SNe we have detected, 
three have $ I_{814} \le 26$ mag, five have $I_{814} \le 27$ mag,
and one has $ I_{814}\sim28.4~$ mag.
SN 1996cl in MS1054.4$-$0321 is a 
spectroscopically verified cluster event, and SN 1999gv in Abell 2218
 is also most 
likely associated with a cluster galaxy. Thus, our upper limits on
the detected field (i.e., non-cluster)
 SN counts, is one event with $ I_{814} \le 26 $
and three events with $ I_{814} \le 27 $. 
These observational results can be compared directly with theoretical
predictions for the number of SNe expected in deep WFPC2 images.

The data in our sample were obtained with a variety of filters and 
exposure times. Converting our results to absolute number counts (i.e.,
SNe per unit sky area, visible at some moment 
to a given flux limit) is further
complicated by a dependence on poorly known parameters for core-collapse
SNe. Instead, we take the published theoretical predictions, fold them through
our observational parameters, apply the model assumptions self-consistently,
and derive the expected numbers of SNe in
our particular sample. We then test the models by direct
comparison of the expected numbers to the observations.

\subsubsection{Detection efficiencies}

We have carried out simulations 
in order to estimate our SN detection efficiencies.
Several tens of artificial point sources (``fake'' SNe) with known flux
and the proper amount of Poisson noise were blindly added to the data for 
each field. The fake SNe were distributed 
such that the probability that 
a galaxy hosts a fake SN is proportional to the galaxy brightness.
The separation of a SN from the galaxy centroid was chosen from a Gaussian
distribution with $\sigma$ equal to 0.85 times 
the galaxy's half-light radius. Thus, the fake-SN 
distribution follows the galactic light. Although in reality the SNe
should follow the galactic {\it luminosity}, flux is a reasonable surrogate;
the cluster galaxy population in a given field is all at a similar distance,
and most of the field galaxies are likely also in a limited redshift range.
Our procedure is also conservative, in the sense that it places the fake 
SNe on too-high galactic backgrounds, leading to an underestimate of the
true efficiency, since
cosmological surface-brightness dimming 
increases the contrast between a SN and its host-galaxy background at high
redshifts. In principle, one might place some fake SNe 
at blank sky positions, to represent SNe in host 
galaxies that are below the detection limit. We find that our 
detection efficiency for such objects is significantly higher than for 
objects placed on galactic backgrounds. However, having no concrete
estimate for the fraction of "hostless" SNe from the total 
population, and in line with our conservative approach,
we did not include such events in the final simulations.
 
 The simulated data were reduced and searched for SNe, like
the real data, and the fraction of recovered objects noted.
Examples of the efficiency curves for the shallowest (Abell 1689)
and deepest (CL1604+4304) data sets are given in Table 3, and the 
relevant efficiency values for all fields are shown in 
Table 4.

\begin{table}
\caption {Sample Detection-Efficiency Values} \label {eta table}
\vspace{0.2cm}
\begin{minipage}{70mm}
\begin{tabular}{ccc}
\hline
Magnitude & \multicolumn {2} {c} {Efficiency}\\ 
$I_{814}$ & Abell 1689   & CL1604+4304   \\    
\hline
23        & 0.95    &  1.0 \\
24        & 0.90    &  1.0 \\
25        & 0.80    &  1.0 \\
26        & 0.45    &  0.85 \\
27        & 0       &  0.60 \\
28        & 0       &  0.50 \\
29        & 0       &  0 \\
\hline
\end{tabular}
\end{minipage}
\end{table}

Six of the nine clusters we
discuss were observed through the F814W filter, which resembles the
Kron--Cousins $I$ filter. 
Model predictions for SN number counts
in deep {\it HST} fields have been  
calculated for the $I$ band (DF; S2000; see
\S3.3.3), and for these fields the comparison is straightforward.
However, for three
of the cluster fields, the duplicate observations were
obtained through the F555W (Johnson $V$), F606W (``wide $V$''), and
F702W (``wide $R$'') filters.

A quantitative comparison of the 
observed $V$-band or $R$-band number counts with the expected $I$-band 
counts requires knowledge of the characteristic colours of the 
high$-z$ SN population. While this information is fairly well known
for type-Ia SNe
(Perlmutter et al. 1999, Schmidt et al. 1998), beyond 25--26~mag, the counts
are expected to be dominated by core-collapse
SNe (e.g., DF; S2000). Broad-band photometry and spectroscopy 
of core-collapse SNe with $z > 0.5$, as well as UV spectra of local
core-collapse SNe, for deriving K-corrections,
are scarce (DF; Gilliland et al. 1999; S2000). 
As the
colours of SNe depend strongly on the SN redshift, type, age,
and extinction, the colour terms are highly model
dependent. The relative fractions of the various sub-types of
core-collapse SNe (II-P, II-L, IIn, Ib, and Ic) are observationally unknown
at high-z, and must be assumed based on sparse data from local observations.
With these limitation in mind, we pursue two alternative
analyses.
 
First, following DF and S2000, we assume that the SN population
probed by the data is dominated by type-II SNe at 
$z \sim 0.7$, about half of which are type II-P events. 
(The actual fraction of the various SN sub-types is controversial; 
we merely make the same assumptions as DF and S2000 in
order to compare their models and our observations self-consistently.)
Based on UV spectra of
local type-II SNe, DF assume that type II-P SNe show a strong UV deficit 
as they enter the plateau phase, some 20 days after maximum light, while
type-II-L and -IIn SNe do not have this effect. At high $z$, the type II-P
SNe will therefore have red $V-I$ and $R-I$ colours. 
We estimate that about half of the type II-P SNe are observed close to 
maximum light, when $V-I$ and $R-I$ are close to zero 
(Fig. 10 of DF), and apply no colour correction to the efficiency
for this fraction of the population. For the remaining
type II-P SNe, 
observed at rest-frame ages $t \geq 20$ days, we adopt the 
colours calculated by DF, based on the SN models by Eastman et al. (1994).
Converting the colours from Fig. 10 of DF to the magnitude system we 
use, we obtain $R_{702}-I_{814}=0.75$ and $V_{555}-I_{814} \geq 2$. Since the 
F606W filter extends further into the red than F555W,
and has significantly higher sensitivity, 
we assume $V_{606}-I_{814}=1.5$. The  
detection efficiency, $\eta$, that applies to the half of the type II-P SN 
population with ages $t \geq 20$ days (1/4 of the total SN population)
is the efficiency found for the observed (non-$I_{814}$) band, at 
a magnitude obtained by adding the colour term. 
For example, in the Abell 2218
data set, the corrected efficiency at $I_{814}=26.4$~mag is 
$[0.75 \times \eta (V_{606} = 26.4) ] + [0.25 \times \eta (V_{606} = 26.4 + 1.5) ]
= [0.75 \times 0.63] + [0.25 \times 0] = 0.47$.

Alternatively, 
in order to compare our observed SN number counts with the predictions in
the least model-dependent way, we consider only 
the sub-sample of six fields observed through the F814W filter,
avoiding the problem of colour corrections altogether. This approach leads
to slightly smaller number statistics, 
but the conclusions, presented below, remain unchanged.   

In all but one of the data sets we have analyzed, the interval between
epochs is ten months or longer, ample time for even the slowest SNe to 
vary significantly, even when time dilation at $z\sim 1$ is taken into 
account. The two epochs we compare in the case of RXJ0848.6+4453, however, 
are separated by only 44 days, equivalent to 22 rest-frame days for a $z=1$ 
SN. Slowly-varying SNe IIn and SNe II-P during their plateau phase could 
become hard to detect in this data set. To correct for this possible 
bias, we again assume that about 1/2 of the high$-z$ SNe we expect to discover
are SNe II-P. These are expected to be in the slowly varying 
phase of their light curve for about 1/3 of the time they are 
observable. We therefore estimate that 1/6 of the expected high$-z$
SNe would not be detected, and apply a correction of 5/6 to the
core-collapse SN detection efficiency in this data set. One
also needs to correct for the possibility that SNe observed on the rise 
in the first epoch would have approximately the same magnitude (within
our observational error) when observed in the declining phase 44 days
later, and thus avoid detection. Using simulated variable objects superposed
on the actual images, we find that $I=26$ objects can vary by at most
0.4 mag and still avoid detection. Applying this to typical light-curves
for the various SN types, we estimate the about 15 per cent of the entire 
SN population might be lost. We therefore apply a further correction
to our efficiency, with the final correction factor being $(5/6) \times
0.85 = 0.7$ .  

\subsubsection{Comparison with theoretical predictions}

S2000 have
considered in detail the exact observational route we have taken --
a search for SNe in
deep, duplicate {\it HST} observations of galaxy cluster fields. Using
models for the expected SN rates, and taking into account various 
models for the lensing effects of the clusters, S2000 have calculated the
number of SNe expected when comparing two images of the
same cluster from different epochs. They find that the SN number counts
depend weakly on the lensing and star-formation models
they use (Table 1 of S2000). 
The background
SN number counts are weakly affected by lensing
because the added depth given by the magnification is countered by
the reduced effective area that is searched, due to that same magnification
(Porciani \& Madau 1999; S2000). 
We can therefore adopt representative values from S2000
for the number of SNe expected per
WFPC2 field for a two-epoch visit. Converting the Vega-based magnitudes
of S2000 (M. Sullivan, private communication)  to AB magnitudes, their
predicted counts are $N_{p} = 1.5$ SNe per WFPC2 field to 
$I_{814}\le26.4$ mag, and $N_{p} = 3$ SNe to 
$I_{814}\le27.4$ mag, in a bi-directional search (i.e., including SNe
discovered in both epochs). 
Note that these numbers are for field 
SNe only, as S2000 exclude cluster events from their 
calculations. S2000 assume 5.71 arcmin$^2$ for the area
of a WFPC2 field, while the correct value for the WF CCDs only
is 4.69 arcmin$^2$, 
so we scale their predictions by 0.82.

In similar work, DF have also calculated
the numbers and properties of high$-z$ SNe. 
Although they address mainly the expected number
of SNe detectable with the Next Generation Space Telescope, 
in their Table 2
they also give the expected number of SNe in an
{\it HST} WFPC2 field to $I_{800}=27$~mag,
equivalent to $I_{814}=26.96$ mag. They predict
a total of 0.5 SN per WFPC2 field per visit. We therefore adopt an
expected number of $N_{p}=1$ SN for every two-visit WFPC2 field reaching
$I_{814}=27$~mag. DF assume 5.0 arcmin$^2$ for the area
of a WFPC2 field, while the value for the WF CCDs only
is 4.69 arcmin$^2$, so we scale their prediction by 0.938.

In our calculations below, we will assume that $1/2$ of the
SNe with  $I_{814} \le 27 $~mag
have $I_{814} \le 26$~mag, based on the results of S2000.
We note that DF calculated the expected SN number counts
in blank (non-cluster) fields, but because of the weak effect of cluster
lensing on total background counts, these predictions can
also be compared to our observations. 
 
Given a predicted SN surface density on the sky, we can calculate the 
expected number of SNe in the observed sample. 
For each data set, we determine the effective area $S$
useful for SN search. This is just the sum of the areas that overlap
the area of a reference epoch. For example,
 if useful data sets are available for two epochs
with complete overlap, the effective area is just twice the illuminated
area of the WF chips,
or 9.38 arcmin$^{2}$. In cases where one of the epochs cannot
be searched, and was used only as a reference,
its contribution to the effective area is zero.
The predicted SN count per magnitude interval, $dm$, per observed
cluster field is thus
the surface density per magnitude interval, $n_p(m) dm$, 
times the fraction of a field
effectively probed for SNe, times the
search efficiency $\eta(m)$, with the latter modified by a colour
term for data sets that were not observed in the $I$ band
(see \S3.3.2):
\begin{equation}
n(m) dm = n_{p}(m) \times \frac{S}{9.38 {\rm arcmin}^2} \times \eta(m)~dm.
\label{eq1}
\end{equation}  
The total SN counts expected per field to a given limiting magnitude can be
obtained by integrating Eq. 1 over $m$. S2000 and DF list their results for
the surface density of SNe, $N_p(<m)$, 
integrated  up to some limiting magnitude,
 rather than the differential surface density $n_p(m)dm$. However, 
since the SN counts are a steeply rising function of magnitude, we
can obtain a good approximation of the total by replacing the integral
with the sum of two terms, for $I_{814}<26.4~(26)$~mag and for $26.4\le I_{814}
\le 27.4~(26\le I_{814} \le 27)$~mag, according to the cumulative numbers 
given by S2000 (DF), and weighted by $\eta(26.4)~[\eta(26)]$ and 
$\eta(27.4)~[\eta(27)]$, respectively. The effective areas, efficiencies, 
and predicted SN numbers for each cluster field are given in Table 4. 

\begin{table*}
\caption {Predicted Field SN Counts} \label {exp table}
\vspace{0.2cm}
\begin{centering}
\begin{minipage}{140mm}
\begin{tabular}{cccccccccc}
\hline
Cluster & Effective area & \multicolumn {4} {c} {Efficiency}& \multicolumn {4} {c} {Expected$^{c}$ SN number ($N$)} \\
        & ($S$)& \multicolumn {4} {c} {($\eta$)} & S2000 & S2000 & DF & DF  \\ 
        & (arcmin$^{2}$)& $I=26$ & $I=26.4$ & $I=27$ & $I=27.4$ & $I \le 26.4$ & $I \le27.4$ & $I \le 26$ & $I \le 27$  \\ \hline 
Abell 2218      & 9.38 & 0.48$^a$ & 0.47$^a$ & 0.46$^a$ & 0        & 0.58  & 0.58 & 0.23 & 0.44\\ 
Abell 1689      & 7.81 & 0.45     & 0.29     & 0        & 0        & 0.30  & 0.30 & 0.18 & 0.18\\ 
AC 114          & 2.56 & 0.69$^a$ & 0.62$^a$ & 0.5$^a$  & 0.41$^a$ & 0.21  & 0.35 & 0.09 & 0.15\\
MS1512.4+3647   & 4.60 & 0.62$^a$ & 0.53$^a$ & 0.37$^a$ & 0        & 0.32  & 0.32 & 0.14 & 0.23\\ 
MS2053.7$-$0449 & 4.38 & 0.77     & 0.70     & 0        & 0        & 0.40  & 0.40 & 0.17 & 0.17\\
MS1054.4$-$0321 & 9.38 & 0.84     & 0.76     & 0.65     & 0        & 0.94  & 0.94 & 0.39 & 0.70\\ 
CL1604+4304     & 7.94 & 0.85     & 0.75     & 0.60     & 0.56     & 0.78  & 1.37 & 0.34 & 0.58\\
RXJ0848.6+4453  & 9.38 & 0.42$^b$ & 0.31$^b$ & 0        & 0        & 0.38  & 0.38 & 0.20 & 0.20\\
CL1645+46       & 6.84 & 0.82     & 0.72     & 0.51     & 0        & 0.65  & 0.65 & 0.28 & 0.45\\
\hline 
\end{tabular} 

Notes:\\ 
$^a$ Corrected with color term to convert to effective
$I$-band efficiency
(\S3.3.2).\\
$^b$ Corrected to account for loss of slowly-fading SNe in
data with short time interval between epochs (\S3.3.2). There
are no offsets between sub-frames of this data set, making the
verification of faint $(I_{814} < 26.5)$ point sources impossible.
The efficiency for $(I_{814} > 26.5)$ is therefore set to zero.\\
$^c$ The predictions of S2000 and DF, as given here, assume 
different cosmologies. For inter-comparison, the predictions
of S2000 should be scaled down by a factor of 0.7  (see text).\\

\end{minipage}
\end{centering}
\end{table*}

Summing over all fields, we obtain the total predicted SN numbers
in the sample according to the models of S2000 and DF.
These predictions are compared with our observations in Table~5.
 
\begin{table*}
\caption {Observed vs. Predicted Field SN Counts} \label {comp table}
\vspace{0.2cm}
\begin{centering}
\begin{tabular}{ccccccccc}
\hline
$I_{814}$ & \multicolumn {2} {c} {Observed (total)} & \multicolumn {2}
 {c} {Observed (non-cluster)}& \multicolumn {2} {c} 
{Expected (S2000)}& \multicolumn {2} {c} {Expected (DF)} \\
mag & all bands & $I_{814}$ & all bands & $I_{814}$ & all bands & $I_{814}$ 
& all bands & $I_{814}$ \\
\hline
$\le26.4$ & 4 & 3 & 2 & 2 & 4.56  & 3.45 & - & - \\
$\le27.4$ & 5 & 4 & 3 & 3 & 5.28  & 4.03 & - & -  \\
$\le26  $ & 3 & 2 & 1 & 1 &   -   &   -  & 2.01 & 1.55 \\
$\le27  $ & 5 & 4 & 3 & 3 &   -   &   -  & 3.09 & 2.27 \\
\hline 
\end{tabular} 
\end{centering}
\end{table*}

Looking at Table~5, one sees that the counts predicted by both DF 
and S2000 are formally consistent with the observations. 
We note, however, that
the predictions of S2000 are always higher than the observed values.
Moreover, the observed numbers given in Table 5 are upper limits,
since, however unlikely, some of the apparent SNe we discuss may not be 
SNe (\S~3.2). Also, the ``non-cluster'' SN sample
includes objects (SNe 1996cp and 1996cq)
for which the redshift is uncertain, and hence may actually be cluster events. 
Finally, (\S~3.3.2) our adopted efficiency values are likely underestimates,
leading to too-low predictions. 
For instance, if SN 1996cp is, in fact, a cluster 
event ($\S~3.1.2$), the difference between the prediction of S2000
and the data becomes statistically significant.

It is therefore instructive to investigate the sources
of the differences 
between the predictions of S2000 and DF. 
The two groups assume very similar
star-formation histories for the Universe (e.g., compare Figure 1 of DF
with Figure 2 of S2000). Each group assumes somewhat different forms
for the delay functions of SNe-Ia (i.e., the SN-Ia rate as a function
of time after an initial burst of star formation), but at these magnitudes, 
Ia's are a small fraction, 15--25 per cent, of all SNe, so this has little
effect on the total counts.
For the UV spectrum of type-II-P SNe during their plateau phase, 
DF use a truncated blackbody model, while S2000 use an empirical spectrum,
but their details are very similar, and in any case, this
is of little consequence, since at $I<27$~mag the median redshift of the SNe 
is $z\sim 0.7$, where the $I$ band probes rest-frame $\sim 4800$~\AA.
The $\Omega_{m}=0.3$, $\Omega_{\Lambda}=0.7$ 
cosmology assumed by S2000 leads only to a
$\sim30$ per cent increase in the total SN number to $I<27$~mag, over that obtained
using DF's assumed $\Omega_{m}=1$ cosmology (P. Nugent, private communication).

The bulk of the discrepancy between the predictions of S2000 and DF  
stems from two subtle differences in the treatment of type-II SNe, for 
which S2000 predict $\sim 2$ times more counts than DF.
First, S2000 use a larger
value for the dispersion of peak absolute magnitudes 
of type II-L SNe ($\sigma = 1.3$ 
mag), derived from the full sample of local type II-L SNe discussed by 
Miller \& Branch (1990), and including the ``overluminous'' II-L SNe 1979C and 
1980K. DF use a smaller dispersion ($\sigma=0.5$) derived from a 
sub-sample excluding the two overluminous SNe, 
but retain the (brighter) peak 
luminosity derived from the full II-L SN sample. 
The large dispersion chosen by S2000 results in a high-luminosity 
tail of type II-L SNe, a large number of which will be visible
 in a flux-limited
SN search. Second, S2000 adopted
 a constant dust extinction value for all type-II SNe, of 
$A_{V}=0.45$ mag, equivalent to $A_{B}=0.66$ mag. DF used a minimum 
extinction value of $A_{B}=0.32$ mag, lower than the one used by S2000, but
modified it according to a random orientation of the host galaxy.
This produced
a large extinction for nearly edge-on hosts, with as many as 
$40$ per cent of the core-collapse SNe suffering extinctions that are greater
than 1~mag (T. Dahl\'en, private communication). This again, 
makes DF's SNe fainter than those of S2000, and the total predicted
numbers lower.

Returning to our observational results,
it appears that the combined assumptions of
DF for the type-II SN peak magnitudes, dispersions, and dust extinctions
may be a better description of the high$-z$ SN population than those
used by S2000. Of course, other variations on these assumptions can be
devised, which will produce the same predicted counts. Furthermore, 
some of the assumptions that are common to the models of both DF and S2000
could also be wrong, but counterbalanced by wrong SN-II parameters.
With improved knowledge of the relative fraction of different SN types,
their luminosity functions, and their spectra, the observed SN counts will
be able to constrain some of the other parameters, such as the 
extinction of core-collapse SNe at high redshift. 

\subsection{Cluster supernova-Ia rates}

\subsubsection{Observed clusters and cluster supernovae}

In this section, we deal with the type-Ia 
SNe in the clusters themselves. As opposed to the
preceding discussion of the field SN counts,
which are dominated by type-II's,
the properties of SNe-Ia are homogeneous and well characterised.
As a result, we can translate the observed numbers into absolute
SN-Ia rates in rich clusters. We separate the clusters
into two subsamples, a low-$z$ sample 
consisting of four clusters at $0.18\le z \le 0.37$, and a high-$z$ sample of
three clusters at $0.83 \le z \le 1.27$.
We exclude from the calculation the cluster
MS2053.7$-$0449, which is at an intermediate redshift ($z=0.58$), and whose 
data are relatively shallow. The existence of the candidate cluster 
CL1645+46 at $z> 1$ was postulated by Jones et al. (1997), 
following the detection in this direction of
a Sunyaev--Zeldovich decrement, as well as 
a $z=3.8$, $198$--arcsec separation quasar pair, PC1643+4631A\&B, 
which was suggested to be the lensed images of a single source. 
 However, Kneissl, Sunyaev, 
\& White (1998) argue that this cluster must be at $z > 2.8$, based 
on an improved upper limit on the X-ray emission. 
Since the observations in our sample cannot detect SNe at such high 
redshifts, this field is automatically excluded 
from our calculations of cluster SN rates.

The area of each cluster covered by the observations
varies within our sample, with
only the cluster core typically covered for the low$-z$ clusters,
but also some of the cluster outskirts for the high$-z$ clusters.
 We therefore limit our calculations of the SN rates to 
a metric radius of $500 h_{50}^{-1}$ kpc around the cluster centres. 
This region is
small enough that significant parts of it are covered by the WFPC2 images
available for the low$-z$ clusters, but large enough to include 
much of the clusters' luminosity. SN 1996cp, which falls slightly outside
this arbitrary region, is excluded from the calculations.      

Among the SNe we have found within the core regions,
two SNe (1999gv and 1996cl) are most probably type-Ia's in their
respective clusters, with one occurring in our low-$z$ cluster sample
and one in our high-$z$ sample. The high-$z$ sample also includes
SN 1996cq, which does not have a
host with known redshift, and so may or may not be a cluster SN-Ia. 
If 1996cq belongs to the $z=0.83$ cluster in whose field it was found, it
is likely a Ia as well, since most core-collapse SNe would be as bright
as this event only briefly at this redshift. We will therefore assume
we have detected one cluster SN-Ia in the low-$z$ sample, and 
one or two cluster SNe-Ia in the high-$z$ sample.
The 95 per cent confidence ranges on the Poisson expectation values for the
number of events in each sample, $N$, are then
$0.07<N<4.75$ at low $z$, and $0.07<N<6.3$ at high $z$, 
where in the latter case we have
taken the union of the 95 per cent probability 
ranges for observing $i\le 1$
and $i\ge 2$ events. We propagate these uncertainties into our rate
calculations, below.

\subsubsection{Rate calculation}

The SN rate per stellar luminosity unit
 in a sample of clusters is obtained by calculating
\begin{equation}
R=\frac{N}{\sum_{i} \Delta t_{i}~f_{i}~L_{Bi}}.
\end{equation}
As detailed below,
$N$ is the observed total number of cluster SNe detected, 
$\Delta t_{i}$ is the effective visibility time of a SN-Ia at 
the cluster redshift,
$f_{i}$ denotes the part of the rest-frame $B$-band
luminosity of a cluster, $L_{Bi}$, that is probed by our SN search,
 and the summation is over all observed cluster
epochs that are useful for a SN search.
By convention, $R$ is often expressed in SNu units, where
1 SNu = 1 SN century$^{-1} (10^{10} L_{B\odot})^{-1}$.

The effective visibility time (sometimes called the ``control time'')
 for a given epoch, $\Delta t_i$, is the period during which a SN-Ia is 
above the detection threshold of the data set, weighted by the 
detection efficiency as a function of magnitude, $\eta(m)$, during this period:
\begin{equation}
\Delta t=\int_{m_{peak, z}}^{m_{lim}} 
\eta(m)(\frac{dm}{dt})^{-1}dm. 
\end{equation}
 Here, $m(t)$ is the SN light curve in the observed band, 
$m_{peak, z}$ is the maximum-light apparent magnitude of a SN 
at the redshift
of the cluster, and $m_{lim}$ is the 
limiting magnitude of the observation ($\eta=0$ for $m>m_{lim}$).
SN-Ia light curves in the specific WFPC2 filters and redshifts 
appropriate to our cluster sample were kindly provided by P. Nugent,
based on the data and calculations in Nugent et al. (2001).
To allow numerical integration of Eq.~3, $\eta(m)$, which was estimated
at discrete values of $m$ in our detection simulations (\S3.3.2), was
linearly interpolated to a finer grid. The effective visibility times
of the sample clusters are listed in Table~5.
Note that, due to cosmological time dilation,
a given time interval in the observer's frame corresponds
to a time shorter by $1+z$ in the rest frame of a cluster at $z$.
However, each SN
light curve is also stretched by $1+z$, and hence is detectable for 
a time that is $1+z$ longer. Due to this cancellation, 
$1+z$ time dilation does not
figure in $N$, the total number of SNe detected.   
The rest-frame SN rate will therefore be proportional to $N$ divided
by the rest-frame effective visibility time, as calculated by 
using, in Eq. 3, SN-Ia light curves without a cosmological $1+z$ stretch
in their time evolution.

As is well documented (e.g., 
Hamuy et al. 1996; Riess et al. 1998; Perlmutter et al. 1997), 
SNe-Ia are not a perfectly uniform 
population, but rather have a non-zero 
range of peak magnitudes and decline rates,
with more luminous SNe fading more slowly and vice versa. Since we 
are finding of order one or less SNe per cluster, this heterogeneity 
in the light curves of SNe-Ia introduces an uncertainty in the 
effective visibility time. For example, in a given cluster, a faint,
quickly declining, SN would have a shorter-than-average
 visibility time, and one
needs to account for the
probability that such a SN is the one that was detected or missed in
that cluster. To this end, we have created a family of SN-Ia light curves
with ``stretch'' factors and correlated peak magnitudes. Based on
the distribution observed by Perlmutter et al. (1999), we assume
SNe-Ia stretch factors between 0.8 and 1.2, with a 
corresponding range in peak absolute magnitudes of $-19.5<M_B<-19.34$.
For each of the observed clusters, we then draw a light curve from the
distribution and find the effective visibility time. The SN rate
for the cluster sample is then calculated. This is repeated many times
in a Monte--Carlo process, to gauge the effect of the light-curve 
non-uniformity on the derived SN rates. We find that the 95 per cent range
of the effect on the final rates
is about $\pm 7$ per cent for the low-$z$ subsample, $\pm 13$ per cent for the high-$z$ 
subsample, and $\pm 10$ per cent for the entire sample.
Peculiar SNe-Ia (brighter, such as SN 1991T, or fainter, 
such as SN 1991bg; see, e.g,
Li et al. 2001, and references within) are observed to be 
rare at high$-z$ (Li et al. 2001), and so will have a 
negligible effect on our derived rates.

The fraction, $f_{i}$, of the luminosity probed at a given epoch
is calculated  by assuming that, within the $500 h_{50}^{-1}$ kpc
radius we consider, the surface brightness profile has a constant,
 $50 h_{50}^{-1}$ kpc core, and then falls as the projected 
radius to the $-1$ power. This is equivalent to assuming the 
light follows a cored isothermal mass distribution 
(e.g., Tyson \& Fischer 1995). By 
integrating over the cluster area with useful coverage in the WFPC2
images, we obtain the fraction of the luminosity within 
$500 h_{50}^{-1}$ kpc that was 
actually searched for SNe. This fraction has a maximum value
of $f=1$ for a cluster in which both available epochs are useful for
SN search, and the overlap area includes the entire 
$500 h_{50}^{-1}$-kpc-radius
region. Typical values of $f_i$ for our sample are $0.45$ for the
low$-z$ sub-sample, and $0.75$ for the high$-z$ clusters.
The values of $f_i$, multiplied by the useful number of epochs, for each
cluster is given in Table~5.
Reducing the adopted core radius by a factor of 5 changes the 
resulting SN rates only by about 5 per cent. Assuming a surface brightness profile
 that falls (unrealistically) like the square of the radius 
lowers the SN rate by about 30 per cent.    
  
Normalizing the SN rate to the stellar luminosity
 requires knowledge of the cluster
luminosities in the rest-frame $B$-band. We have found in the literature
luminosities for several of the clusters in our sample. Where 
necessary, the published values have been corrected to our assumed 
cosmology. 
We have taken the $V$-band luminosities for clusters with $z\sim0.3$ and
the $I$-band luminosities for clusters with $z\sim0.85$ as reasonable 
representations of the rest frame $B$-band luminosities. Our low$-z$ 
sample consists of four rich clusters.
Abell 2218 and AC114, which have published optical luminosities, 
are quite similar to Abell~1689 and MS1512.4+3647, which do not 
(this is also apparent from the {\it HST} images used in this work),
so we have assumed a value of $L_B=2.5\pm0.5 \times 10^{12} h_{50}^{-2} 
L_{B\odot}$ for the two clusters lacking published data,
based on the Abell 2218 and AC114 measurements. 
Our high$-z$ sample includes two optically rich systems (MS1054.4$-$0321 and 
CL1604+4304) shown to be at least as rich as the Coma cluster, 
with a mean value of $L_{B}(<500 h_{50}^{-1} \rm{kpc}) \sim 5.5 \times 10^{12}~h_{50}^{-2} L_{B\odot}$. MS1054.4$-$0321 does not have a published 
error estimate for its luminosity, and we assume an uncertainty of 
$\pm 1.5 \times 10^{12}~h_{50}^{-2} L_{B\odot}$.
The third high-$z$ cluster, RXJ0848.6+4453 at $z=1.27$, 
presents a difficulty,
 as it is possibly poorer than the other two.
However, few examples of clusters at such high $z$ exist,
and estimates of typical luminosities have not been made.
We therefore adopt a 
plausible luminosity value for this cluster of half the mean luminosity 
of the other two, with a $1\sigma$ uncertainty of one half the adopted
value. This gives a probable range for the luminosity between zero
(equivalent to dropping the cluster from the sample) 
and the high luminosity of the other two clusters.

To assess the effect of the uncertainties of the cluster luminosities 
on the rate calculation, we have 
included in the Monte--Carlo simulation described above, in addition
to a light-curve draw, a luminosity draw for each cluster. 
Each cluster's luminosity was drawn from a Gaussian distribution with
mean and $\sigma$ as given in Table 6.
If the uncertainty in the visibility times is not included in the 
simulation (i.e., only a standard SN-Ia light curve is assumed) the effect 
of the luminosity uncertainty on the final rates is $\pm 15$ per cent at low $z$, $+55$ per cent, $-30$ per cent at high $z$, and $+38$ per cent, $-20$ per cent for the entire sample (95 per cent confidence ranges).
 The large uncertainty in the luminosity of RXJ0848.6+445
has a minor effect because, at its high redshift, the visibility time
of SNe-Ia is short, giving it low weight in the calculation.
The uncertainties in the luminosities of the clusters clearly dominate
over the uncertainties in the SN-Ia light-curve shape in the final SN-rate
errors. 

\begin{table*}
\caption {Cluster Luminosities and SN-Ia Visibility Times} \label {lum table}
\vspace{0.2cm}
\begin{centering}
\begin{minipage}{140mm}
\begin{tabular}{cccc}
\hline
Cluster & $L_B^a$& $f_i\times $& $\Delta t_i^c$\\
        & [$10^{12} h_{50}^{-2}L_{B\odot}$]& No. epochs$^b$& [rest-frame days]\\
\hline
Abell 2218    & $2.5 \pm 0.5~~^{d} $ & 1.00 & 219 \\
Abell 1689    & $2.5 \pm 0.5~~^{h}         $ & 0.80 & 200 \\
AC114         & $2.5 \pm 0.25~~^{e}$ & 0.99 & 208 \\
MS1512.4+3647 & $2.5 \pm 0.5~~^{h}         $ & 0.94 &  147 \\ 
MS1054.4$-$0321 & $5.5\pm 1.5~~^{f}$& 1.75 & 101 \\
CL1604+4304   & $6.5\pm 2~~^{g}$      & 1.00 & 105 \\
RXJ0848.6+4453& $3\pm 1.5~~^{h}         $ & 1.50 &  19 \\
\hline 
\end{tabular} 

Notes:\\ 
$^a$ Rest-frame $B$-band luminosity within $R<500 h_{50}^{-1}$ kpc, and adopted
$1\sigma$ error, assuming  $\Omega_m=0.3$, $\Omega_{\Lambda}=0.7$.\\
$^b$ Fraction $f_i$ of the luminosity surveyed for SNe, times the number 
of epochs used for search.\\
$^c$ Effective visibility time, in days,
 of SNe-Ia at the cluster redshift, in cluster rest-frame.\\
$^d$ Squires et al. 1996\\
$^e$ Natarajan et al. 1998\\
$^f$ Hoekstra, Franx \& Kuijken 2000\\
$^g$ Postman, Lubin, \& Oke 2001\\
$^h$ No measurement available in the literature, see text.\\

\end{minipage}
\end{centering}
\end{table*}
 
With this input,
we obtain from Eq. 2 a rest-frame cluster SN-Ia
rate of $R=0.20^{+0.84}_{-0.19} h_{50}^{2}$ SNu at $0.18<z<0.37$ and 
$R=0.41^{+1.23}_{-0.39}h_{50}^{2}$ SNu at $0.83<z<1.27$, where
the errors include the 95 per cent confidence Poisson upper and lower 
limits, as well as the 68 per cent ranges due to the combined
light-curve and luminosity uncertainties discussed above.
Although the errors are large, we see no indication
for any extreme evolution of the SN-Ia rate between the two 
redshift bins. We therefore 
also calculate a SN-Ia rate based on the two or three SNe
in the combined cluster sample, with 
$0.18 < z < 1.27$, and a mean redshift (weighted by the rest-frame
effective visibility time of each cluster) of $\bar z=0.41$. We find
$R = 0.30^{+0.58}_{-0.28}h_{50}^{2}$ SNu, with the errors having 
the same meaning as above.   

\begin{table}
\caption {SN-Ia Rates} \label {rates table}
\vspace{0.2cm}
\begin{centering}
\begin{minipage}{70mm}
\begin{tabular}{cccc}
\hline
Environment & Redshift & SN Rate$^a$ & Reference\\ 
            &          & $h_{50}^{2}$[SNu] &  \\    
\hline
Cluster (low$-z$)  & 0.25   & $0.20^{+0.30}_{-0.13}$ & This work  \\
Cluster (high$-z$) & 0.90   & $0.41^{+0.47}_{-0.21}$ &  \\
Cluster (all)   & 0.41   & $0.30^{+0.21}_{-0.11}$ &  \\
\hline
E-S0   & local  &  $0.07\pm0.03$ & Cappellaro \\
S0a-Sb & local  &  $0.09\pm0.03$ & et al. (1997)\\
Sbc-Sd & local  &  $0.11\pm0.04$ & "\\
Cluster& 0.06   &  $0.11^{+0.06}_{-0.07}$ & Reiss (2000) \\
Field  & 0.11   &  $0.10^{+0.07}_{-0.07}$ & "\\
Field  & 0.10   &  $0.11^{+0.09}_{-0.05}$ & Hardin et al. 2000 \\
Field  & 0.38   &  $0.20^{+0.14}_{-0.09}$ & Pain et al. 1997\\
Field  & 0.55   &  $0.16^{+0.03}_{-0.02}$ & Fabbro et al. 2000\\
\hline
\end{tabular}

Note: $^a$ All quoted errors are $1\sigma$. 
\end{minipage}
\end{centering}
\end{table}

\subsubsection{Comparison to models and to other measurements}

Table 7 summarises our rate measurements and compares them 
to those available in the literature. 
In local elliptical galaxies (which are the dominant population in the
central regions of rich
clusters), Cappellaro et al. (1997) find  
$R = 0.067\pm 0.03 h_{50}^{2}$~SNu. Among distant field galaxies,
 measured rates are 
 $R = 0.20^{+0.14}_{-0.09}h_{50}^{2}$~SNu 
at $\bar z=0.38$ (Pain et al.
1997) and $R = 0.16^{+0.03}_{-0.02}h_{50}^{2}$~SNu at $\bar z=0.55$ (Pain et al.
2002; see Fabbro et al. 2000), where the quoted errors are $1\sigma$.
The SN-Ia rates we have measured in the central regions of rich clusters
 at $0.18<z<1.27$ are not markedly different from the rates in the local
elliptical population, or in the general field population at similar
redshifts.  

As mentioned in \S1, a major motivation for measuring cluster SN rates
is the possibility of addressing directly the puzzle of the large
mass of iron and the excess entropy observed in clusters. For
example, using 
gas-dynamical models for ICM enrichment by SNe, Brighenti \&
Mathews (1998) find that the cluster SN-Ia rate must be 
$> 1.2 h_{50}^{2}$~SNu  today and
$> 2.4 h_{50}^{2}$~SNu at $z=1$ to explain production
of most of the iron in the ICM with SNe-Ia. These values are above
our 95 per cent upper limits on the cluster SN-Ia rates.
We note, however,
that the uncertainty in our
measured rates is dominated by the large Poisson errors resulting from small 
number statistics. Even a slightly larger cluster SN sample could 
more critically test the validity of the SN-Ia ICM enrichment scenario. 
   
\section{Conclusions}

We have conducted a search for SNe using data from
deep observations of galaxy clusters from the
{\it HST} archive, and have detected six point-like transient events. 
Based on their observed characteristics and the known properties
of different classes of variable objects, we have argued that most 
or all of these events are SNe. The redshifts and types of the likely
host galaxies are available in the literature for five 
events, demonstrating the benefits of observing well-studied fields. 
 
Folding in the observational parameters of our search, we have
compared the numbers and magnitudes of the likely non-cluster (i.e.,
background or foreground) SNe we have found to published predictions of
SN number counts. Such predictions depend strongly on a number of 
poorly known ingredients, and in particular the luminosity function
and extinction of core-collapse SNe. The observed counts from our survey
provide some initial constraints on these models. The observed counts,
which are comparable to the predictions of DF, but somewhat 
lower than those of S2000, suggest that at $z\sim 0.7$, core-collapse SNe
are relatively faint, due to either extinction or to a luminosity 
distribution without a high-luminosity tail. However, being based on
small SN numbers, these results should be treated with caution,
and larger surveys, yielding significantly more events, are definitely
called for.  

We have used the candidate type-Ia SNe in the clusters
to estimate
the SN-Ia rate in the central $500 h_{50}^{-1}$~kpc of
medium-redshift $(0.18 \le z \le 0.37)$ and high-redshift
 $(0.83\le z \le 1.27)$ 
cluster sub-samples.
The measured rates
are consistent with SN-Ia measurements in field environments,
both locally and at high redshift. Our results argue against an environmental
dependence of the SN-Ia rate. 
Specifically, the observed low SN rate
is in conflict with
a scenario where SNe-Ia are the primary source of the
large amounts of iron seen in the ICMs of massive clusters.
 
This work shows that useful
limits on high$-z$ SN rates and counts can be derived from repeated,
deep {\it HST} imaging, even without any follow-up observations.  
The current dominant source of uncertainty is the Poisson error
due to small number statistics. Next in importance are the lack of 
identification for some of the SNe, and the uncertainties in cluster
luminosities. Obviously, similar analysis 
of a  larger data set would yield SN rates and counts with higher precision, 
and planned observations (rather than an archival search) would enable
some followup work that would reduce the systematic errors. 
Such an expanded program would also improve the prospects of discovering
and utilising gravitationally lensed SNe, as outlined in \S1.

Excluding the search for SNe in the HDF by Gilliland et al. (1999), 
our search is the deepest
SN survey conducted, and covers about 6 times the effective area 
of the HDF search. We find, on average, roughly one SN per deep 
($I_{814} > 26$~mag) 
duplicate WFPC2 cluster field, with about half of the SNe in the
clusters and half in the field. At these magnitudes, 
there is a high density of SNe on the sky, detectable in significant
numbers even in the small field of view of the {\it HST} cameras.
Finding these SNe and characterising them, even if only in an
incomplete and statistical way, can provide input and constraints
to many current issues in astronomy and cosmology. If you are about to
carry out deep imaging, whether from space or from the ground, do not do it
all at once! Numerous SNe are out there, and can be found simply by 
splitting the observations into two or more well-separated epochs.

\section*{Acknowledgments}

We thank P. Nugent for kindly calculating and providing
SN light-curve templates before publication, and helping to resolve the
discrepancies in published SN count predictions.
We also thank B. Milvang-Jensen and A. Arag{\'o}n-Salamanca for
kindly providing redshifts for galaxies in MS1054.4-0321
prior to publication.   
N. Brosch, T. Dahl\'en,  P. van-Dokkum,  P. Madau, E. Ofek,
I. Smail, M. Sullivan, and S. Zucker are thanked
for valuable data and advice,
and the anonymous referee for numerous useful suggestions.
This research has made use of 
the NASA/IPAC Extragalactic Database (NED), which is operated by the 
Jet Propulsion Laboratory, California Institute of Technology, under 
contract with the National Aeronautics and Space Administration.  
This work was supported by the Israel Science Foundation -- the Jack Adler
Foundation for Space Research, grant 63/01-1.

\end{document}